# Corroborative evidences of $TV^{\gamma}$-scaling of the α-relaxation originating from the primitive relaxation/JG β relaxation

K. L. Ngai[1]*, M. Paluch[2,3]*

[1]*CNR-IPCF, Dipartimento di Fisica, Università di Pisa, Largo B. Pontecorvo 3, I-56127, Pisa, Italy*

[2]*Institute of Physics, University of Silesia, Uniwersytecka 4, 40-007 Katowice, Poland*

[3]*Silesian Center for Education and Interdisciplinary Research, 75 Pulku Piechoty 1, 41-500 Chorzow, Poland*

*to whom the correspondence should be addressed: ngaikia@gmail.com ; marian.paluch@us.edu.pl

**Abstract**

Successful thermodynamic scaling of the structural α-relaxation time $\tau_\alpha$ or transport coefficients of glass-forming liquids determined at various temperatures *T* and pressures *P* means the data conform to a single function of the product variable $TV^{\gamma}$, where *V* is the specific volume and γ is a material specific constant. In the past two decades we have witnessed successful $TV^{\gamma}$-scaling in many molecular, polymeric, and even metallic glass-formers, and γ is related to the slope of the repulsive part of the intermolecular potential. The advances made indicate $TV^{\gamma}$-scaling is an important aspect of the dynamic and thermodynamic properties of glass-formers. In this paper we show the origin of $TV^{\gamma}$-scaling is not from the structural α-relaxation time $\tau_\alpha$. Instead it comes from its precursor, the Johari-Goldstein β-relaxation or the primitive relaxation of the Coupling Model and their relaxation times $\tau_\beta$ or $\tau_0$ respectively. It is remarkable that all relaxation times, $\tau_\alpha$ $\tau_\beta$ and $\tau_0$ are functions of $TV^{\gamma}$ with the same γ, as well as the fractional exponent $\beta_K$ of the Kohlrausch correlation function of the structural α-relaxation. We arrive at this conclusion convincingly based on corroborative evidences from a number of experiments and molecular dynamics simulations performed on a wide variety of glass-formers and in conjunction with consistency with the predictions of the Coupling Model.

1. Introduction

The dependence on temperature $T$ and specific volume $V$ of thermodynamic and dynamic properties of glass-formers is exactly known if the interaction is approximated by a spherically symmetric, pairwise additive, repulsive potential,

$$U(r) \propto (\sigma/r)^{3\gamma}. \tag{1}$$

Here $r$ is the intermolecular separation, and σ and γ are material-specific constants related, respectively, to the molecular size and the steepness of the potential. Governed by the potential, the properties are functions of the variable $TV^{\gamma}$.[1-4] Attractive forces are present in liquids as well, such as that represented by an extra term in the generalized Lennard-Jones (L-J) potential of the form,

$$U(r) = \frac{\varepsilon}{(p-q)}[q(\sigma/r)^p - p(\sigma/r)^q], \tag{2}$$

with the powers $p$ and $q$ of the repulsive and attractive parts of the potential.

Since the last decade, it has been shown that the relaxation times of the structural α-relaxation, $\tau_\alpha(T,P)$, of many glass-formers determined at various combinations of temperature $T$ and pressure $P$ can be superimposed to conform to a unique function of $TV^\gamma$,

$$\tau_\alpha(T,V) = \tau_\alpha(TV^\gamma) \qquad (3)$$

This was first shown for OTP with $\gamma = 4$.[5-7] Others [8-24] have generalized Eq.(3) to a wide range of glass-forming liquids and polymers with values of γ within the range $0.13 \leq \gamma \leq 8.5$ by dielectric spectroscopy. There are confirmations from measurements using light scattering, [21], molecular dynamics simulations [23-27], static ambient-pressure quantities [28,29], and viscosity [30-32]. In polymers, $TV^\gamma$-scaling was found for both $\tau_\alpha(T,P)$ and the relaxation time $\tau_n(T,P)$ of the chain normal modes with the same γ by simulations [23] and by dielectric relaxation [33-35], i.e.

$$\tau_n(T,V) = \tau_n(TV^\gamma), \qquad (4)$$

Relaxation times of several different correlation functions of the molten salt $2Ca(NO_3)_2 \cdot 3KNO_3$ (CKN) obey $TV^\gamma$-scaling with the same γ [18,36] In an epoxy resin, $TV^\gamma$-scaling was successfully applied to both $\tau_\alpha(T,P)$ and the Johari-Goldstein (JG) relaxation time $\tau_\beta(T,P)$ with the same γ, [18] i.e.

$$\tau_\beta(T,V) \approx \tau_\beta(TV^\gamma), \qquad (5)$$

The functions, $\tau_\alpha(TV^\gamma)$, $\tau_n(TV^\gamma)$, and $\tau_\beta(TV^\gamma)$, are all different, albeit the scaling exponent γ is the same. The '≈' sign is used in (5) because of the following reason. The primitive relaxation time $\tau_0(T,P)$ of the Coupling Model (CM) [37-40] is exactly a function of $TV^\gamma$ with the same γ as $\tau_\alpha(TV^\gamma)$ in Eq.(1), i.e.

$$\tau_0(T,V) = \tau_0(TV^\gamma), \qquad (6)$$

because of the exact CM relation,

$$\tau_\alpha(T,P) = [t_c^{-n}\tau_0(T,P)]^{1/(1-n)}, \qquad (7)$$

or

$$\tau_\alpha(TV^\gamma) = [t_c^{-n}\tau_0(TV^\gamma)]^{1/(1-n)}, \qquad (8)$$

where $t_c$ is about 1 to 2 ps for van der Waals glass-formers, and the coupling parameter $n$ is the complement of the fractional exponent $\beta_K=(1-n)$ of the Kohlrausch stretch exponential correlation function,

$$\varphi(t) = \exp[-(t/\tau_\alpha)^{1-n}], \qquad (9)$$

The Fourier transform of which is used to fit the frequency dispersion of the α-relaxation. It is important to recognize the general and experimentally established fact [41,42] that the frequency dispersion of the α-relaxation, or $\beta_K=(1-n)$, is invariant to changes in $T$ and $V$ as long as $\tau_\alpha(TV^\gamma)$ is maintained constant. Thus, $\beta_K(TV^\gamma)$ or $n(TV^\gamma)$ is a well-defined function of $TV^\gamma$ with the same γ as $\tau_\alpha(TV^\gamma)$. With this point clarified, Eq.(3) rewritten in full is given by

$$\tau_\alpha(TV^\gamma) = [t_c^{-n(TV^\gamma)}\tau_0(TV^\gamma)]^{1/[1-n(TV^\gamma)]} \qquad (10)$$

However, the value of $\tau_\beta(T,P)$ determined experimentally by some arbitrary fitting procedure by workers can only be expected approximately equal to $\tau_0(T,P)$,[17,40-42] i.e.

$$\tau_\beta(P,T) \approx \tau_0(P,T), \text{ and } \tau_\beta(TV^\gamma) \approx \tau_0(TV^\gamma), \qquad (11)$$

and thus the approximate equality sign in Eq.(5).

The approximate equality (11), $\tau_\beta(P,T) \approx \tau_0(P,T)$, have been verified in many glass-formers, mostly at ambient pressure where experimental data of $\tau_\beta(T,P)$ are plentiful and $\tau_0(T,P)$ are readily calculated by Eq.(7). [40] This is an ancillary property supporting the CM, worth taken into account in further development of this paper to follow.

From Eqs.(10) and (11), we arrive at the approximate relation,

$$\tau_\alpha(TV^\gamma) \approx [t_c^{-n(TV^\gamma)} \tau_\beta(TV^\gamma)]^{1/[1-n(TV^\gamma)]} \tag{12}$$

Thus, if the experimentally deduced value of $\tau_\beta(T,V)$ is approximately a function of $TV^\gamma$ with the same $\gamma$ as $\tau_\alpha(TV^\gamma)$, the prediction of the CM is in effect verified because $\tau_0(T,V)$ is exactly a function of $TV^\gamma$ with the same $\gamma$ as $\tau_\alpha(TV^\gamma)$.

From the account of various aspects of $TV^\gamma$-scaling of $\tau_\alpha(T,P)$ given above, it is not only generally obeyed in non-associated glass-formers, but also it is fundamentally important because of its putative connection to the molecular potential [1-4,16,17]. The simultaneous $TV^\gamma$-scaling of $\tau_\alpha(T,P)$ together with other relaxation times including $\tau_\beta(T,V)$ with the same $\gamma$ but different functions adds another intrigue to the origin. The question that arises from this is which relaxation does the $TV^\gamma$-scaling originates from? Although $TV^\gamma$-scaling was mostly performed on the structural α-relaxation, in this paper we provide corroborative evidences from experiments and simulations to conclude that $TV^\gamma$-scaling originates from the JG β-relaxation time $\tau_\beta(T,P)$ or the related primitive relaxation time $\tau_0(T,P)$ of the CM[37-40]. Having arrived at this conclusion, we show that it can explain other experimental facts related to $TV^\gamma$-scaling in molecular and polymeric glass-formers, which in return bolsters our conclusion.

## 2. Glass-formers having unresolved JG β-relaxation

There are many molecular glass-formers in which the JG β-relaxation is unresolved because $\tau_\beta(T,P)$ is not much shorter than $\tau_\alpha(T,P)$, and instead appears as an wing-like excess loss over the fit by the Fourier transform of the Kohlrausch function to the dielectric α-loss peak on its high frequency side [41-44]. Examples include cresolphthaleindimethyl ether (KDE), phenylphthalein-dimethyl ether (PDE), propylene carbonate (PC), chlorinated biphenyl (PCB62), phenyl salicylate (salol), dibutyl phthalate (DBP), 1,1′-di(4-methoxy-5-methylphenyl) cyclohexane (BMMPC) [41,42], and many more. The α-loss peak in these materials is usually narrow and the value of $n$ from the KWW fit is small and the $\tau_0(T,P)$ calculated by Eq.(7) has the corresponding primitive frequency, $f_0(T,P) \equiv 1/2\pi\tau_0(T,P)$, located within the excess wing. From the approximate relation (11) it is clear that the excess wing is the unresolved JG β-relaxation [40,43,44,45]. The example from PDE is shown in the inset of Fig.1, where in the main figure we present the $TV^\gamma$-scaling of $\tau_\alpha(T,P)$ from dielectric measurements with γ=4.4.

Any given value of $\tau_\alpha$ or loss peak frequency $f_\alpha$ at ambient pressure can be maintained the same upon increases in pressure $P$ by raising temperature. In dielectric studies of KDE, PDE, PC, PCB62, salol, DBP, BMPC, BMMPC [41,42] and other molecules [40], different combinations of $P$ and $T$ had been chosen to maintain $f_\alpha$ or $\tau_\alpha$ constant. Remarkably, there is no change in the shape or frequency dispersion of the entire dielectric loss composed of the α-loss peak and the excess wing on its high frequency flank (see inset of Fig.1). This temperature-pressure superpositioning of the entire frequency dispersion at constant $f_\alpha$ is valid for different choices of $f_\alpha$. This general result implies that the ratio $\tau_\alpha(T,P)/\tau_\beta(T,P)$ as well as $\beta_K(T,P)=[1-n(T,P)]$ is invariant to changes of $P$ and $T$ at constant $\tau_\alpha(T,P)$ or constant $\tau_\beta(T,P)$ even though no attempt is made to resolve the two relaxations. Since $TV^\gamma$-scaling was demonstrated to holds for $\tau_\alpha$ from experimental data of $\tau_\alpha(T,P)$ in these exemplary molecular glass-formers, it follows from the invariance of $\tau_\alpha(T,P)/\tau_\beta(T,P)$ or $\tau_\alpha(T,P)/\tau_0(T,P)$, and $\beta_K(T,P)=[1-n(T,P)]$ that $TV^\gamma$-scaling works also for $\tau_\beta(T,P)$ or $\tau_0(T,P)$, and $\beta_K(TV^\gamma)=[1-n(TV^\gamma)]$ with the same γ. The $TV^\gamma$-scaled function of $\beta_K(T,P)$, $\beta_K(TV^\gamma)=[1-n(TV^\gamma)]$, is shown in Fig.2. The functions $\tau_\beta(TV^\gamma)$ and $\tau_0(TV^\gamma)$ are different from $\tau_\alpha(TV^\gamma)$ because[18] it follows from Eqs.(10) or (12) that the ratio $\tau_\alpha(TV^\gamma)/\tau_0(TV^\gamma)$ or $\tau_\alpha(TV^\gamma)/\tau_\beta(TV^\gamma)$ is not constant but increases with $\tau_\alpha(TV^\gamma)$.

The co-invariance of $\tau_\alpha(T,P)$, $\tau_\beta(T,P)$, $\tau_0(T,P)$, and $\beta_K(T,P)$ to variations of $P$ and $T$ is a remarkable property[17,40-43], which is the key leading to $\tau_\alpha(TV^\gamma)$, $\tau_\beta(TV^\gamma)$, $\tau_0(TV^\gamma)$, and $\beta_K(TV^\gamma)$ are all functions of $TV^\gamma$ with the one and the same γ, and the bedrock of many thermodynamic scaling properties to be discussed.

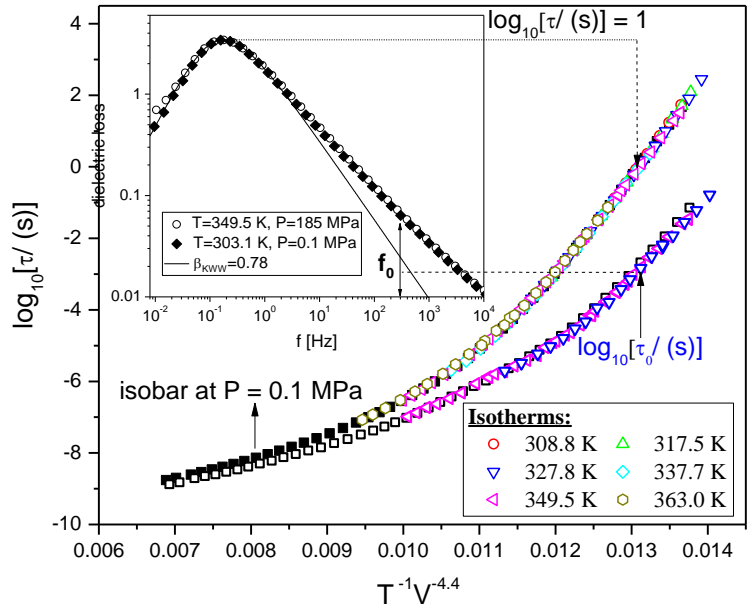

**Figure 1.** The inset shows example of perfect superpositioning of the dielectric loss spectra of a van der Waals glass-former, PDE, at different combinations of $P$ and $T$ for a constant $\tau_\alpha$. The primitive frequency, $f_0=1/2\pi\tau_0$, is indicated by the arrow. Consequently, $TV^\gamma$ scalings of $\tau_\alpha$ as well as $\tau_0$ hold as shown in the main figure.

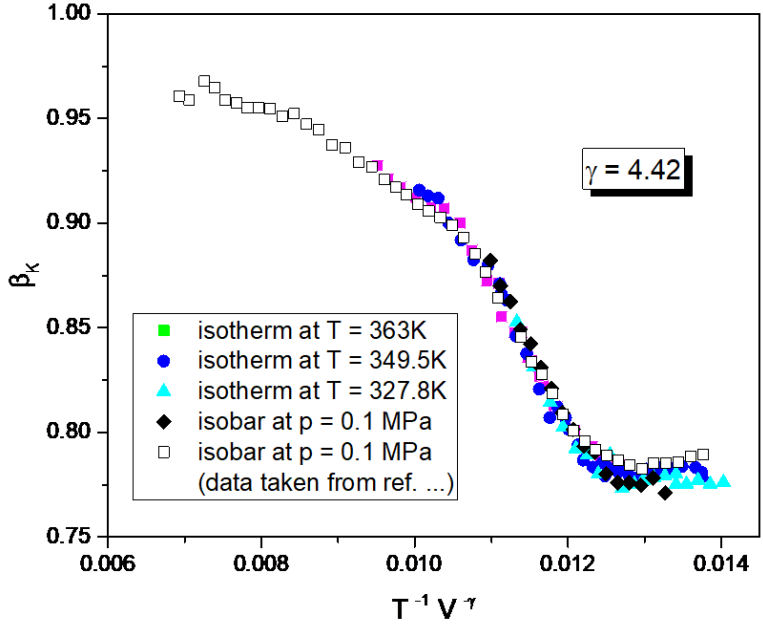

**Figure 2.** $TV^\gamma$-scaled $\beta_K(TV^\gamma)=[1-n(TV^\gamma)]$ as a function of $TV^\gamma$ with the same $\gamma$ as $\tau_\alpha(TV^\gamma)$ and $\tau_0(TV^\gamma)$ or $\tau_\beta(TV^\gamma)$ in Fig.1.

## 3. Glass-formers having resolved JG β-relaxation

### 3.1 DGEBA

Our first attempt to show simultaneous $TV^\gamma$-scaling for both $\tau_\beta(T,P)$ and $\tau_\alpha(T,P)$ was made in diglycidyl ether of bisphenol-A (DGEBA) with $M_w$=380 g/mol (also known as EPON 828) obtained by the isobaric dielectric relaxation measurements at ambient and elevated pressures by changing temperature, and at several fixed temperatures by changing pressure. $\tau_\alpha(TV^\gamma)$ and $\tau_\beta(TV^\gamma)$ are found to be functions of $TV^\gamma$, with the same $\gamma$=3.5.[18] Although most of the data of the scaled $\tau_\beta(TV^\gamma)$ are in the glassy state, there are only a few in the liquid state. This is because the JG β-relaxation of DGEBA in the dielectric loss spectra cannot be unambiguously resolved if $\tau_\alpha(TV^\gamma)$ becomes much shorter than 100 s. Nevertheless, the limited amount of resolved JG β-relaxation data do support simultaneous $TV^\gamma$-scaling of $\tau_\alpha(TV^\gamma)$ and $\tau_\beta(TV^\gamma)$.

Furthermore, the primitive relaxation times $\tau_0(TV^\gamma)$ at several state points in the liquid state have been calculated by Eq.(10) with $\beta_K\equiv(1-n)=0.52$ obtained previously by fitting the frequency dispersion of the α-relaxation to the Fourier transform of the Kohlrausch function. The calculated $\tau_0(TV^\gamma)$ are in agreement with $\tau_\beta(TV^\gamma)$, verifying the ancillary property of the JG β-relaxation mentioned before. This guarantees $\tau_\beta(TV^\gamma)$ obey thermodynamic scaling with the same γ as of $\tau_\alpha(TV^\gamma)$ because $\tau_\beta(TV^\gamma) \approx \tau_0(TV^\gamma)$ and $\tau_0(TV^\gamma)$ automatically does the same in view of Eq.(10). All the results discussed above are contained in Fig.S1 in the Supporting Information.

## 3.2 Methylated derivative of ketoprofen

Methylated derivative of ketoprofen is incapable of forming local hydrogen bonding structures. Its dielectric spectra shows the excess loss over the Kohlrausch fit, which becomes a shoulder at higher frequencies contributed by a partly resolved secondary relaxation [46]. The entire loss, peak and shoulder, is invariant to changes of $P$ and $T$ (see Fig.S2(a) in the Supporting Information). This is evidence of a JG β-relaxation supported by its pressure sensitivity, and also the consistency of its relaxation time $\tau_\beta(T,P)$ with the calculated $\tau_0(T,P)$.[44,47,48] Most interesting is that $\tau_\alpha(T,P)$ can be density scaled simultaneously together with $\tau_\beta(T,P)$ as function of $TV^\gamma$ with same value of the γ exponent, and for sure also the calculated $\tau_0(T,P)$. This clear observation of simultaneous density scaling of $\tau_\alpha(T,P)$ and $\tau_\beta(T,P)$ of a resolved JG β-relaxation with the same γ in a supercooled liquid (see Fig.S2(b) in the Supporting Information) throws another support for $\tau_\beta(T,P)$ or $\tau_0(T,P)$ as the originator of $TV^\gamma$-scaling.

Since the JG β-relaxation was not fully resolved, fitting the shoulder as an additive contribution to the α-loss peak as done in Ref.[46] is inconsistent with the CM's view of evolution of dynamics and non-additivity.[40,48] This inconsistency accounts for about one order of magnitude difference between the calculated $\tau_0(T,P)$ and $\tau_\beta(T,P)$ from the fit at $T=235$ K and $P=0.1$ MPa shown in the inset of Fig.2 in Ref.[45], and between the calculated $\tau_0(TV^\gamma)$ and $\tau_\beta(TV^\gamma)$ from the fit in Fig.6b of Ref.[45]. This does not change the $TV^\gamma$-scaling of the primitive and JG β-relaxation times with the same γ as $\tau_\alpha$, but it has to be pointed out in view of Eq.(11).

### 3.3 Molecular dynamics simulations of a polymer

Bedrov and Smith [49] performed molecular dynamics simulations of a bead-spring model of polymer with interaction between beads governed by the Lennard-Jones potential at various combinations of $P$ and $T$ while keeping $\tau_\alpha(T,P)$ constant. They found the JG β-relaxation at times preceding the α-relaxation in the torsional auto-correlation function. The time dependence of the α-relaxation (or $n$) as well as the ratio $\tau_\alpha(T,P)/\tau_\beta(T,P)$ are invariant too (see Fig.3S(a) in the Supporting Information). The findings of co-invariance of $\tau_\alpha(T,P)$, $\tau_\beta(T,P)$, and $n$ are independent of the choice of the constant $\tau_\alpha(T,P)$ as shown in Fig.S3(b) in the Supporting Information. Although no thermodynamic scaling was performed, but the co-invariance will ensure $\tau_\alpha(T,P)$, $\tau_\beta(T,P)$, and $n(T,P)$ will have the same γ. Therefore, the results from the simulation provide another evidence for $TV^\gamma$-scaling works simultaneously for $\tau_\beta(T,P)$, $\tau_\alpha(T,P)$, and $n(T,P)$ with the same γ.

### 3.4 The pharmaceutical ternidazole

The strongly hydrogen-bonded ternidazole drug (3-(2-methyl-5-nitroimidazol-1-yl)-propan-1-ol, $C_7H_{11}N_3O_3$, and TDZ for short, was studied for its temperature and pressure dependence of the specific volume and dielectric relaxation in both the supercooled liquid and glass states.[50] Two secondary relaxations were detected. The slower one is the JG $\beta$-relaxation because it shift to lower frequencies on elevating pressure[43] in concert with the α-relaxation, and also its relaxation time satisfies Eq.(11) over a range of $P$ and $T$.[40-43] Like DGEBA discussed in subsection 3.2, the JG $\beta$-relaxation is not fully resolved in the liquid state, and appears as an excess loss on the high frequency flank of α-loss peak. Notwithstanding, the frequency dispersion of the entire loss is invariant to changes of $P$ and $T$ at constant α-loss peak frequency (see Fig.S4(a) in the Supporting Information). As said before in subsection 3.1, even without making any fits to resolve the two relaxations and determine $\tau_\beta(T,P)$ and $\tau_\alpha(T,P)$, this property implies the invariance of the ratio $\tau_\alpha(T,P)/\tau_\beta(T,P)$ at any fixed $\tau_\alpha(T,P)$. The latter in turn assures $\tau_\beta(T,P)$ and $\tau_\alpha(T,P)$ will simultaneously obey $TV^\gamma$-scaling with the same γ, as demonstrated in Fig.S4(b) in the Supporting Information. The authors of Ref.50 also find good agreement of the calculated $\tau_0$ from the CM Eqs.(7) or (8) with the observed $\tau_\beta$.

## 4. Direct evidence from short time dynamics

When time becomes sufficiently short to make cooperativity of dynamics vanishing, the α-relaxation is the same as the primitive or the JG β-relaxation. If $TV^\gamma$-scaling of the observed relaxation time or transport coefficient is successful, the result can be taken as direct evidence of $TV^\gamma$-scaling of $\tau_\beta(T,P)$ or $\tau_0(T,P)$.

**(a) Low viscosity data**

Studied by dielectric relaxation, the $P$ and $T$ dependence of $\tau_\alpha(T,P)$ in the archetypal molecular glass-formers discussed in the previous section is usually limited to the longer time range of $\tau_\alpha(T,P)$. Notwithstanding, the dynamics in the short time regime can be inferred from measurements of the viscosity $\eta$ at elevated pressures and temperatures. $TV^\gamma$-scaling of the $\eta(T,V)$ data was successful in many glass-formers and ionic liquids [31-33],

$$\eta(T,V) = \eta(TV^\gamma), \qquad (13)$$

with a scaling parameter $\gamma$ close to that of $\tau_\alpha(TV^\gamma)$ represented by Eq.(3) found by dielectric relaxation measurements. Conforming to $TV^\gamma$-scaling are the viscosities measured in the range

$$10^{-3} \leq \eta(T,V) \leq 1 \text{ Pa.s.} \qquad (14)$$

This range of $\eta(T,V)$ corresponds to dielectric relaxation times in the range

$$10^{-12} \leq \tau(T,V) \leq 10^{-9} \text{ s} \qquad (15)$$

from the Maxwell relation, $\eta(T,V)/G_\infty$, if the infinite-frequency shear modulus $G_\infty$ is $10^9$ Pa as found by mechanical relaxation at lower temperatures for van der Waals. The value of $G_\infty$ can become smaller at high temperatures but not by much.

The α-relaxation of many molecular glass-formers including PC, PDE, salol, and DBP have been characterized by dielectric measurements at high frequencies up to 1 GHz [51,52], by dynamic light scattering measurements of OTP at times of order of 1 ns and shorter [21,53], and by depolarized Rayleigh scattering of OTP, DOP (dioctalphthalate), aroclor, and toluene at times less than 1 ns.[54] The width of the frequency dispersion is either slightly broader or the same as the Debye relaxation or the exponent $\beta_K=(1-n)$ of the Kohlrausch stretch exponential function is equal to or nearly equal to 1. Therefore practically the α-relaxation becomes the primitive relaxation or the JG β-relaxation, a result which also follows from Eq.(7) or Eq.(11). Thus, the $TV^\gamma$-scaling, found for $\eta(T,V)$ in the range given by (14) or for $\tau(T,V)$ in the corresponding range of (15), is actually for the primitive relaxation or the JG β-relaxation. In other words, the successful density scaling of viscosity in the range given by (14) is tantamount to another evidence of density scaling of $\tau_\beta(T,V)$ and $\tau_0(T,V)$ in the range of (15) with the same γ as density scaling of the dielectric $\tau_\alpha(T,V)$ at longer times as shown for dibutylphthalate [30] with γ=3.2 (see Fig.S5 of Supporting Information).

The relaxation times from depolarized light scattering [21] and other characteristic times from neutron scattering [5] of OTP are also short and within the time range of (15). Hence the $TV^\gamma$-scaling of the data in these high frequency experiments is for $\tau_\beta(T,V)$ and $\tau_0(T,V)$, and with the same γ as low frequency data.

**(b) Diffusion data from simulations**

Diffusion coefficient $D$ and the reduced $D^*$ were obtained at different isobaric conditions by molecular dynamics simulations of several Lennard-Jones m-6 liquids, where the power $m$ of the repulsive part of the potential falls within the range, (8≤m≤36).[27] The $D$ and $D^*$ are found to be a unique function of the variable $\rho^\gamma/T$, where ρ is density (see Fig.S6(a) in the Supporting Information). The scaling exponent $\gamma$ is related to the steepness of the repulsive part of $U(r)$, evaluated around the distance of closest approach between particles probed in the supercooled regime. The data of $D$ obtained span roughly 5 decades of change, about three decades of which are at temperatures corresponding to the self intermediate scattering function being exponential [27], i.e. the primitive relaxation. Only 2 decades of $D$ is below the temperature $T_0$,[55] where non-exponential α-relaxation typical of the supercooled regime first becomes apparent upon cooling the liquid. Thus, the data demonstrate $\rho^\gamma/T$-scaling is valid for the exponential primitive/JG relaxation above $T_0$ and the non-exponential α-relaxation below $T_0$ with the same scaling exponent γ.

## (C) Self-intermediate scattering function

Self-intermediate scattering functions $F_s(k^*, t^*)$ of the Lennard-Jones $m$-6 liquids with $m=12$ were obtained by simulations[56] at five densities, $\rho=1.15, 1.20, 1.25, 1.30$, and $1.35$, and from which the reduced relaxation times $\tau^*$ are found to obey $\rho^\gamma/T$-scaling with $\gamma=5.1\pm 0.1$ (see Fig.S6(b) in Supporting Information). The range of $\tau^*$ include the short times regime where the functions has exponential time dependence. Hence the primitive relaxation time is included in the figure and its relaxation time also obey $\rho^\gamma/T$-scaling with the same $\gamma$ as longer $\tau^*$ of the $\alpha$-relaxation. This point is consistent with another simulations of the LJ 12-6 system in Ref.[57] where density scaling is carried out by $E_\infty(\rho)/T$ instead of $\rho^\gamma/T$. Notwithstanding, before the onset of slow or cooperative dynamics, the relaxation with short relaxation times is primitive because its time correlation function is exponential.

In the inset of Fig.S6(b) in Supporting Information, $F_s(k^*, t^*)$ for each of the five densities, and at the respective temperatures corresponding to a fixed value of $\rho^\gamma/T=5.07$ are plotted as a function of reduced time. $F_s(k^*, t^*)$ has essentially the same shape for state points for which $\rho^\gamma/T$ is constant. This result shows not only do the relaxation times superpose as a function of $\rho^\gamma/T$, but also the entire $t$-dependence of the correlation functions is invariant for isomorphic state points. The longer time portion of $F_s(k^*, t^*)$ is the $\alpha$-relaxation, while the short time part of $F_s(k^*, t^*)$ is the primitive relaxation. The same shape of the longer time portion means invariance of the time dependence of the $\alpha$-relaxation $\tau_\alpha$ or its value of $n$, and the latter is also a function of $\rho^\gamma/T$ with the same $\gamma=5.1\pm 0.1$. The invariance of the $t$-dependence of the primitive relaxation implies co-invariance of its relaxation time $\tau_0$ in concert with $\tau_\alpha$ and $n$, a property found in real glass-formers. This result was reinforced in another report of simulations of the dynamics of L-J systems.[57] The contents of Fig.S6(b) together with the inset are tantamount to those in Fig.1 in this paper. In both cases $\rho^{\gamma}/T$-scaling applies to the primitive or exponential relaxation at short relaxation times and the cooperative and non-exponential α-relaxation with long relaxation times.

The Kohlrausch exponent, $\beta_K = (1-n)$, of the self-intermediate scattering function was obtained by fitting the time dependence, and found invariant to thermodynamic conditions at constant $\tau_\alpha(\rho,T)$ or $\tau_0(\rho,T)$.[58] The $\rho^{\gamma}/T$-scaling of $\tau_\alpha(\rho,T)$ is also obeyed by $\beta_K = (1-n)$ exhibited in Fig.S6(c) in the Supporting Information. Once again $\beta_K$ is a function of $\rho^{\gamma}/T$ as in Fig.2.

Moreover, the full $t$-dependence of the four-point dynamic susceptibility, $\chi_4(t)$, and thus the maximum in $\chi_4(t)$, which is proportional to the dynamic correlation volume, are invariant for state points for which the scaling variable $\rho^{\gamma}/T$ is constant.[56]

## 5. $TV^{\gamma}$-scaling originates from the JG β-relaxation/primitive relaxation

The evidences presented in Sections 2-4 have sufficiently shown that the JG β-relaxation time $\tau_\beta(T,V)$ or the primitive relaxation time $\tau_0(T,V)$ also obey $TV^{\gamma}$-scaling like $\tau_\alpha(T,V)$ and with the same γ. In the following we give subsidiary facts to support that the $TV^{\gamma}$-scaling originates from $\tau_\beta(T,V)$ with the function $\tau_\beta(TV^{\gamma})$, or from $\tau_0(T,V)$ with the function $\tau_0(TV^{\gamma})$, and pass it onto $\tau_\alpha(TV^{\gamma})$ in the course of time with the development of many-molecule cooperativity.

### 5.1 From the principle of causality

The simplest argument is to invoke the physical principle of causality[59]. Since the JG β-relaxation from experiment/simulations and the primitive relaxation in the CM transpire at times earlier than the α-relaxation, from causality their $TV^{\gamma}$-scaling property has to be the origin or cause of the $TV^{\gamma}$-scaling of the α-relaxation, viscosity, and diffusion. Thus $TV^{\gamma}$-scaling of $\tau_\alpha(T,V)$ is the result of starting from the basic $\tau_\beta(TV^{\gamma})$ or $\tau_0(TV^{\gamma})$ and incorporating the subsequent slowing-down by cooperativity in the inherent many-molecule dynamics of the α-relaxation. The combined effect is given in the CM by using Eq.(10) or relation (12) where in general $n(TV^\gamma)$ depends only on the relaxation time $\tau_\alpha(TV^\gamma)$, or $\tau_0(TV^\gamma)$ and approximately $\tau_\beta(TV^\gamma)$. The fact that $n(TV^\gamma)$ is uniquely determined at any value of $\tau_\alpha(TV^\gamma)$, and hence is a well-defined function of $TV^\gamma$, comes from the invariance of the frequency dispersion of the α-relaxation to changes of $T$ and $P$ or $V$ while $\tau_\alpha(TV^\gamma)$ is kept constant [40-43]. The results are rendered by Eq.(10), and by the approximate relation (12). In the low viscosity regime of (14) or the short time regime of (15), $n(TV^\gamma)$ has zero value, and (10) and (12) are reduced to $\tau_\alpha(TV^\gamma) = \tau_0(TV^\gamma)$ and $\tau_\alpha(TV^\gamma) \approx \tau_\beta(TV^\gamma)$, i.e. the $TV^\gamma$-scaling of $\tau_\beta(T,V)$ and $\tau_0(T,V)$ are the ones directly observed.

**5.2 From γ determined from slope of the potential at short distance**

**(A) Lennard-Jones *m*-6 liquids**

Another argument for $TV^\gamma$-scaling originating from $\tau_\beta(T,V)$ or $\tau_0(T,V)$ is from the remarkably small value of *r* at which the steepness of the repulsive part of the intermolecular potential $U(r)$ determines the scaling exponent γ. This provides an important clue because it is generally believed that structural relaxation in liquids is governed by $U(r)$. The short distance *r* was obtained by molecular dynamics simulations of several Lennard-Jones *m*-6 liquids, where the power *m* of the repulsive part of the potential falls within the range, (8≤*m*≤36).[27] The diffusion coefficients for all the simulated Lennard-Jones *m*-6 supercooled liquids at various *P* and *T* are found to be a unique function of the variable $\rho^\gamma/T$, where $\rho$ is density and proportional to the reciprocal of *V*. The scaling exponent $\gamma$ is related to the steepness of the repulsive part of $U(r)$, evaluated around the distance of closest approach between particles probed, which is significantly shorter than that at the first peak of the radial distribution functions $g(r)$ between the particles (see Fig.S7 in the Supporting

Information). At such short distance where $\gamma$ is determined, it is hard to make any connection with the dynamically heterogeneous α-relaxation having much larger length-scale. On the other hand, the local primitive relaxation and the JG β-relaxation conceivably are governed by the repulsive part of $U(r)$ at short distance, and they are conferred the $\rho^\gamma/T$ scaling property. The latter is passed on to the α-relaxation at longer length and time scales as conveyed by Eq.(10) and relation (12) of the CM.

**(B) *cis*-1,4-polybutadiene**

Molecular dynamics simulations of cis 1,4-polybutadiene by Tsolou et al. [24] using a Lennard-Jones 12-6 intermolecular potential together with harmonic backbone bonds to characterize the forces between chain segments. Both segmental relaxation time $\tau_\alpha(T,V)$ and normal mode relaxation time $\tau_n(T,V)$ obtained for different $T$ and $V$ superposed by density scaling as functions $\tau_\alpha(TV^\gamma)$ and $\tau_n(TV^\gamma)$ with the same γ=2.8. This value of γ is smaller than 4 expected for the LJ 12-6 potential from 3γ≈12, and is due to the softening influence of the intramolecular bonds as demonstrated in Ref.[30]. A power law $r^{-8.5}$ is found in the repulsive interaction at short distance with the power 8.5 equal to 3γ (see Fig.S8 in the Supporting Information). However, the power law $r^{-8.5}$ was found at distance less than 2.5 Å, whereas the intermolecular pair distribution function, $g(r)$, obtained from the simulations has the first peak at ~ 5 Å, and decreases to zero before 2.5 Å is reached. [60] Again, $TV^\gamma$-scaling originating from $\tau_\beta(T,V)$ and $\tau_0(T,V)$ is supported by the short distance at which γ is determined by the slope of the potential in cis 1,4-polybutadiene.

Worthwhile to mention in passing are two other results from the simulations.[24] One is the time autocorrelation function of the dipole moment shows the nearly exponential decay for $t<t_c$ and changes to the stretched exponential for $t>t_c$. The crossover time $t_c$ is between 1 and 2 ps. This result from simulation verifies the basic premise of the CM [37-40], which leads to the predictions given by Eqs.(3)-(12). Another found by simulations is the invariance in the frequency dispersion of the segmental α-relaxation to variations of *P* and *T* at constant $\tau_\alpha(T,P)$.

## 5.3 Absence of correlation of γ with *n*(TV$^\gamma$) or *m* at $\tau_\alpha$(TV$^\gamma$)=100 s

Quantifying the frequency dispersion of the α-relaxation, the fractional exponent $\beta_K \equiv (1-n)$ of the Kohlrausch function is uniquely defined at any fixed value of $\tau_\alpha(T,V)$ independent of *T* and *V*. Therefore after density scaling the data, *n* is a well-defined function of $TV^\gamma$ as well as $\tau_\alpha$. If $TV^\gamma$-scaling originates from the α-relaxation, a correlation between γ and $n(TV^\gamma)$ at some fixed value of $\tau_\alpha(TV^\gamma)$ can be expected. A test has been made by using most of the published dielectric data of γ and *n* with the choice of $\tau_\alpha(TV^\gamma)=100$ s,[40] but no correlation is found. There is no correlation between γ and the fragility parameter *m* either. The absence of correlation are shown in Figs.S9 and S10 in the Supporting Information with more recent data included. This negative result suggests that γ has no clear connection to the α-relaxation per se. However, as already been mentioned, γ is related to the slope of the intermolecular potential at short distance, and density scaling starts at the primitive/JG relaxation level.

The absence of correlation between γ and $\beta_K \equiv (1-n)$ or *n* is readily understood from the $TV^\gamma$-dependence originating from that of $\tau_0$ or $\tau_{JG}$, and the stronger dependence of $\tau_\alpha$ on $TV^\gamma$ is due to the build-up of many-body dynamics. In the context of the CM, $\tau_\alpha(TV^\gamma)$ is obtained from $\tau_0(TV^\gamma)$ or $\tau_\beta(TV^\gamma)$ by Eq.(10) or Eq.(12) by raising the dependence to the superlinear power of $1/(1-n)$, i.e. $\tau_\alpha(TV^\gamma) \propto [\tau_0(TV^\gamma)]^{1/(1-n)}$. From this relation, it is clear that the nonexponentiality index *n* of the α-relaxation and γ are *independent* parameters and they bear no obvious correlation with each other. Furthermore, *n* depends on anharmonicity determined by the attractive part as demonstrated by molecular dynamics simulation[61], while γ is about 1/3 of the slope of the repulsive part of the potential. The isobaric fragility index $m_P$ (say at ambient pressure) of the α-relaxation obtained from the $T_g/T$ dependence of $\tau_\alpha$. Hence from Eqs.(10) or (12), $m_P$ is determined by both $n$ and $\gamma$, and therefore like $n$, $m_P$ has no correlation with $\gamma$.

## 6. $TV^\gamma$-scaling with the same γ for the bulk and in nano-confinement

A remarkable property of density scaling were found in the comparative study of bulk and nano-confined van der Waals liquids, tetramethyltetraphenyl-trisiloxane (DC704) and polyphenyl ether (5PPE).[62] The α-relaxation time $\tau_\alpha(T,P)$ of the core of the confined material not only obey $TV^\gamma$-scaling like the bulk, but also with the same value of the scaling exponent γ (see Fig.S11 in the Supporting Information). These results have an explanation based on the $TV^\gamma$-scaling originating from $\tau_0(TV^\gamma)$ or $\tau_\beta(TV^\gamma)$, which is unchanged upon nano-confinement. According to Eq.(10) or Eq.(11), $\tau_\alpha(T,P)$ is also a function of $TV^\gamma$ with the same γ, irrespective the material is nano-confined or bulk. The same value of the scaling exponent for bulk and confined liquids is ultimately the consequence of same intermolecular potential.

## 7. $TV^\gamma$-scaling of different dynamic quantities of the same material with the same γ

It is remarkable that density scaling is obeyed for different dynamic variable or quantities of the same material with the same γ. We have considered this property before in the case of CKN[18,36]. Having proven that the $TV^\gamma$-scaling originates from $\tau_0(TV^\gamma)$ of the primitive relaxation, this property can be explained by generalizing Eq.(10) to different dynamic variables or modes μ with different $n_\mu(TV^\gamma)$.

Another case is the normal modes and α-relaxation of polymers [32-34], where the two processes conform to density scaling with the same γ as found by dielectric spectroscopy, and by molecular dynamics simulations of 1,4 polybutadiene mentioned before in Subsection 5.2(B). The same steps as taken in the above for $\tau_\alpha(T,P)$ of the core of the confined material were applied before to explain why the normal mode relaxation time $\tau_n(T,P)$ also obey $TV^\gamma$-scaling with the same γ as $\tau_\alpha(T,P)$,[33,34] i.e. Eq.(4). This was done before in Ref.[35], and further developed in Ref.[63]. Here we point out from experiment[34] that, at fixed value of the relaxation time of $\tau_n(T,P)$ or $\tau_\alpha(T,P)$, the shape of the relaxation peak for either the normal mode or the α-relaxation is constant. Thus the Kohlrausch shape parameter (1- $n_n$) or the coupling parameter $n_n$ of the normal mode is uniquely determined at constant $\tau_n(T,P)$ and is a well-defined function of $TV^\gamma$ as well. Some small deviations of simultaneous $TV^\gamma$-scaling of the two modes with the same γ were found in low and high molecular weights of polyisoprene[64]. A possible explanation of the deviations is the slight change of frequency dispersion of the normal mode on elevating pressure as shown in Fig.10 of ref.63, while there is no change in the case of the segmental α-relaxation.

There is also the MD simulations on the ionic glass-former, $2Ca(NO_3)_2 \cdot 3KNO_3$ (CKN), performed by Ribeiro *et al.* performed at various *T* and *P*.[65] The non-polarizable pairwise potential was given by a Born-Mayer function. Diffusion coefficient, *D*, relaxation time of the intermediate scattering function, $\tau_\alpha$, and $NO_3^-$ anion reorientational time, $\tau_r$, were obtained as a function of *T* and *ρ*. All these three dynamical properties of CKN scale as $\rho^\gamma/T$ with a common value γ = 1.8 ± 0.1.

More recently, the property of $TV^\gamma$-scaling with the same γ for different variables is found by molecular dynamics simulation in two very different materials, and again it can be explained by $TV^\gamma$-scaling originating at the primitive/JG β-relaxation level. One is a ternary metallic glass-forming liquid.[66] The average dynamics, dynamic heterogeneities including the high order dynamic correlation length, and static structure are all well described by thermodynamic scaling with the same scaling exponent γ.[66] The other is molecular dynamics simulations of a polymer melt.[67] The same exponent γ applies in the $TV^\gamma$-scaling of the fast mobility (i.e., the mean square amplitude of the picosecond rattling motion inside the cage), and the much slower structural relaxation and chain reorientation.

The fact of successful $TV^\gamma$-scaling with the same γ for different quantities implies that the relations between these quantities such as their relaxation times are invariant to changes of $P$ and $T$ at any fixed value of $TV^\gamma$. The invariance in their relations is highly nontrivial. Therefore it has to come from some fundamental cause, which we believe is the primitive relaxation being the origin of $TV^\gamma$-scaling of all others.

## 8. Conclusion

From the co-invariance of $\tau_\alpha(T,P)$, $\tau_\beta(T,P) \approx \tau_0(T,P)$, and $n(T,P)$ found in many van der Waals molecular glass-formers, whether the JG β-relaxation is resolved or not, we have demonstrated that the primitive relaxation and the JG β-relaxation times as well as $n$ are functions of $TV^\gamma$, i.e. $\tau_0(TV^\gamma)$, $\tau_\beta(TV^\gamma)$, and $n(TV^\gamma)$, with the same γ as the α-relaxation time $\tau_\alpha(TV^\gamma)$. We have reached the same conclusion from the successful $TV^\gamma$-scaling of the low viscosity $\eta(TV^\gamma)$ and high diffusion coefficient $D(TV^\gamma)$ at high temperatures with the same γ as the dielectric $\tau_\alpha(TV^\gamma)$. This is because the corresponding relaxation times from such $\eta(TV^\gamma)$ and $D(TV^\gamma)$ are so short that they correspond to that of the primitive relaxation/JG β-relaxation.

Since the primitive relaxation/JG β-relaxation is the precursor of the α-relaxation, the principle of causality is applied to assert that $TV^\gamma$-scaling originates from $\tau_0(TV^\gamma)$ or $\tau_\beta(TV^\gamma)$. Moreover, the scaling exponent γ is related to the steepness of the repulsive part of $U(r)$, evaluated around the distance of closest approach between particles probed. This distance is so short that unlikely γ is related to the dynamically heterogeneous α-relaxation of much longer length-scale.

This is supported by the absence of correlation between γ and the Kohlrausch stretch exponent (1-$n$) or the fragility parameter $m$ of the α-relaxation. The absence of correlation is readily rationalized by the CM Eqs.(10) and (12) from the fact that $n$ and γ are separate and independent parameters.

Based on the origin of $TV^\gamma$-scaling is coming from $\tau_0(TV^\gamma)$ or $\tau_\beta(TV^\gamma)$, the experimental fact that the α-relaxation time of nano-confined liquids obeys $TV^\gamma$-scaling with the same γ as the bulk material can be explained. Also explained in the same way is the $TV^\gamma$-scaling with the same γ of different dynamic quantities or variables in CKN, polymers, and metallic glasses.

We mention in passing that exact analogues of the properties considered in the previous sections of non-ionic glass-formers are found in the conductivity relaxation of ionic liquids.[68,69] There is a secondary ionic conductivity relaxation which is related to the primary conductivity relaxation (responsible for d.c. conductivity) in the same way as the JG β-relaxation is related to the structural α-relaxation on pressure and temperature dependence. Hence $TV^\gamma$-scaling of the α-conductivity relaxation also originates from the primitive or the secondary ionic conductivity relaxation.

The corroborating evidences collected in this paper from different materials and effects should be sufficient to conclude in general that the primitive relaxation or the JG β-relaxation is the initiator of the $TV^\gamma$-scaling of glass-formers, which is passed on to the structural α-relaxation in the course of time with the development of the many-body cooperative dynamics. Since $TV^\gamma$-scaling embodies both thermodynamic and dynamic properties, the fundamental importance of the primitive relaxation or the JG β-relaxation in regard to these basic properties of glass-formers in general should not be overlooked in solving the glass transition problem.

**Supporting Information**

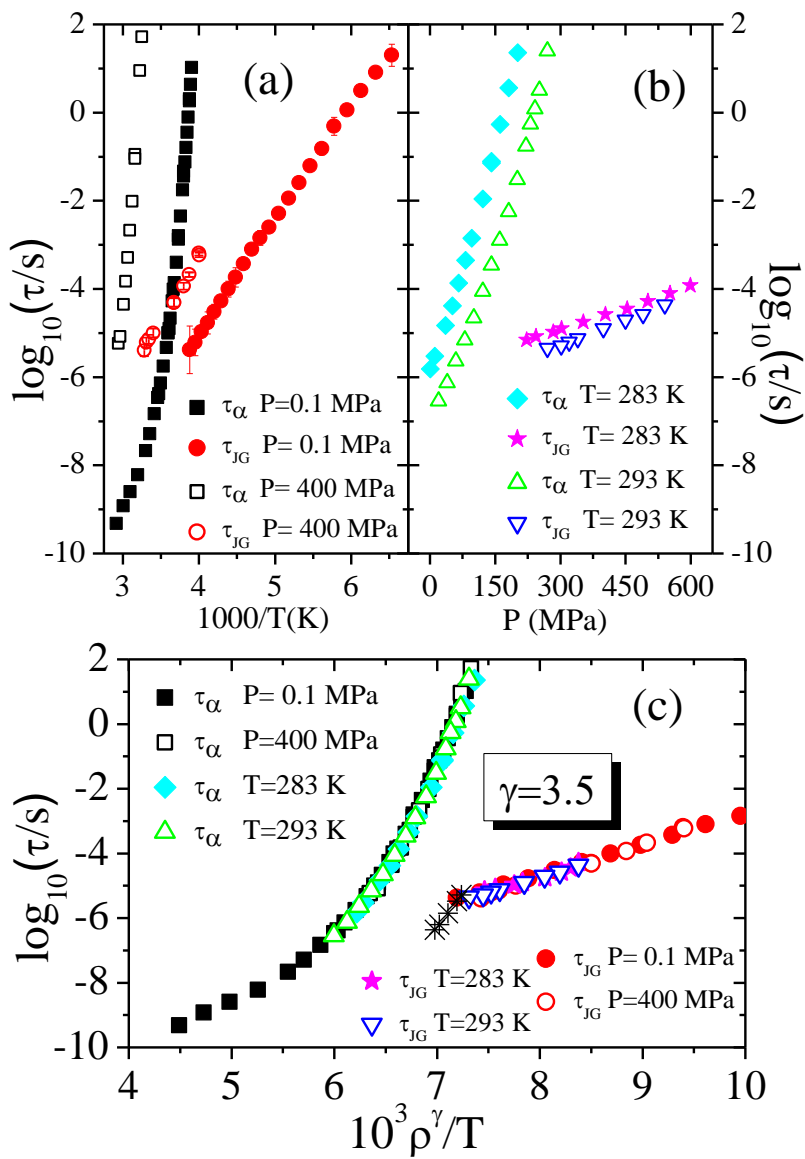

**Figure S1.** Logarithm of characteristic time of dielectric loss maximum of DGEBA (diglycidyl ether of bisphenol-A, $M_w$=380 g/mol, also known as EPON 828) for α-relaxation and JG-relaxation in isobaric condition versus reciprocal temperature (a), in isothermal condition versus pressure (b), and an overall plot of the same data versus $\rho^\gamma/T$ (c). When not shown, error bars are smaller than symbol size. Black asterisks in panel (c) indicates the values for $\log_{10}(\tau_0)$ at several state points calculated by the Coupling Model Eq.8 with the stretch exponent $\beta_K\equiv(1-n)=0.52$ obtained by fitting the frequency dispersion of the α-relaxation by the Fourier transform of the Kohlrausch-Williams-Watts function. Figure reproduced by permission of J.Chem.Phys.

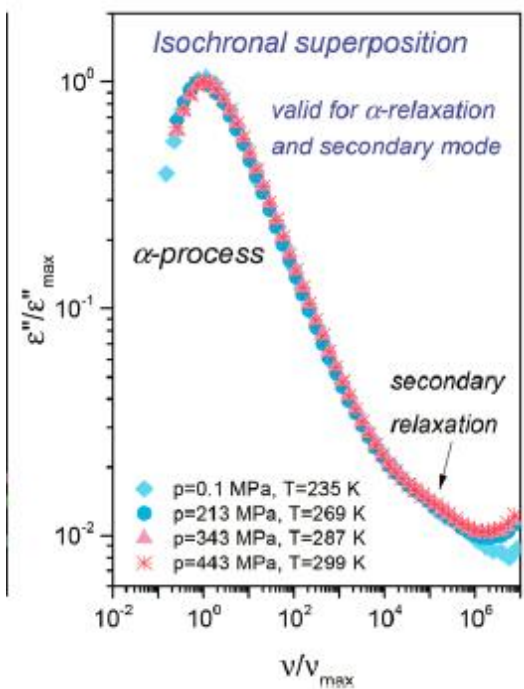

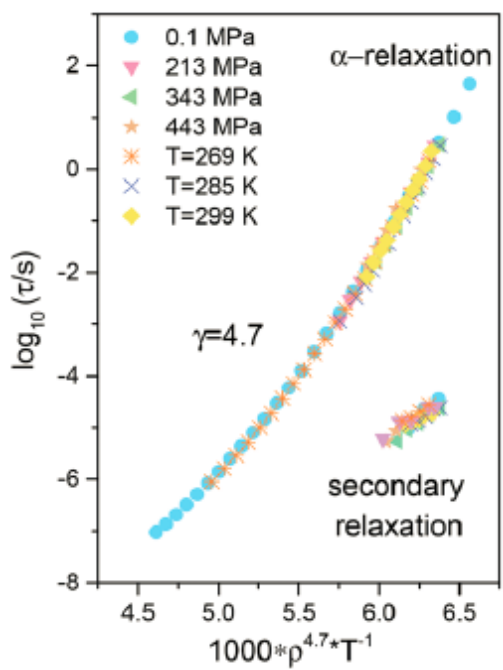

**Figure S2.** (a）Normalized dielectric loss spectra for Me-RS-ketoprofen taken at different (T, P) conditions but with approximately the same peak frequency. (b) α- and β- relaxation times

plotted as a function of 1000/ $T\rho^{-4.7}$ for Me-RS-ketoprofen with γ=4.7. Density data were taken from PVT data. Figures reproduced by permission.

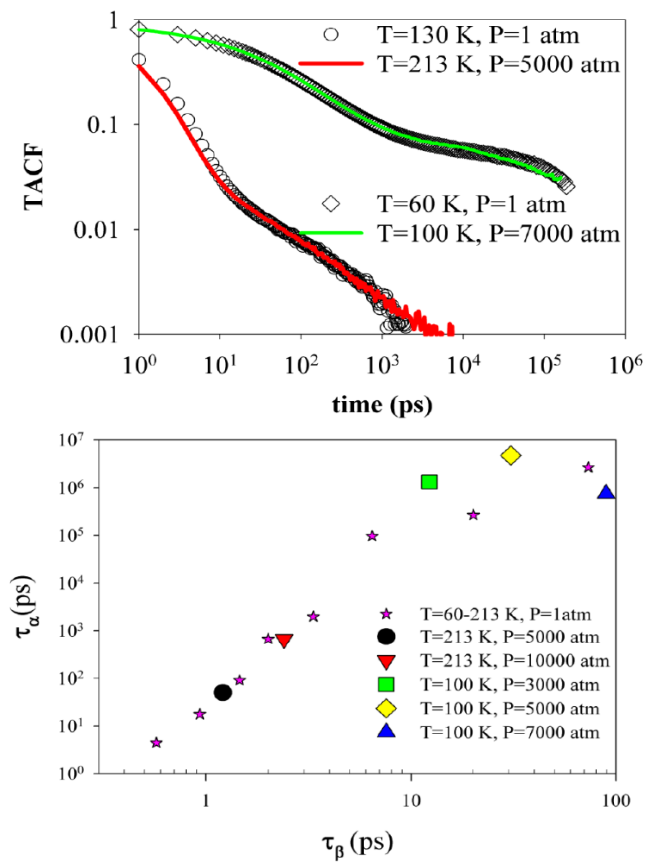

**Figure S3**. (a) TACF obtained from simulations of melts with l=1.5 Å (σ/l=2.7) at different thermodynamic conditions defined by temperature and pressure. (b) Correlation between relaxation times for α- and β-processes as obtained from simulations of melts with l=1.5 Å (σ/l=2.7) at various thermodynamic conditions. Figure reproduced by permission by J.Non-Cryst. Solids.

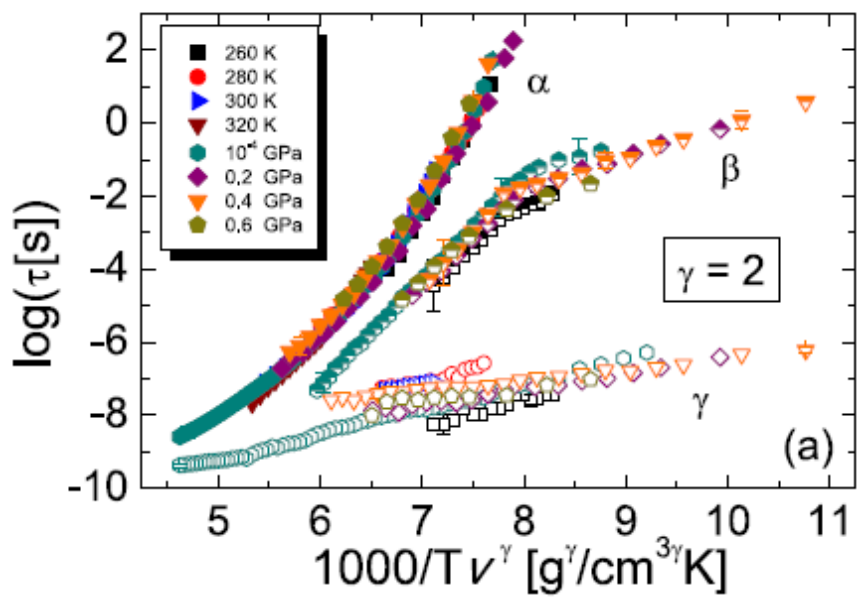

**Figure S4.** Thermodynamic scaling plot of the relaxation times of all three dynamic processes in ternidazole drug, as function of the quantity $1000/TV^{\gamma}$ with $\gamma = 2$ ($V$ is the specific volume calculated for each ($P$, $T$) pair by means of the experimental equation of state. Representative error bars are indicated. Figure reproduced by permission.

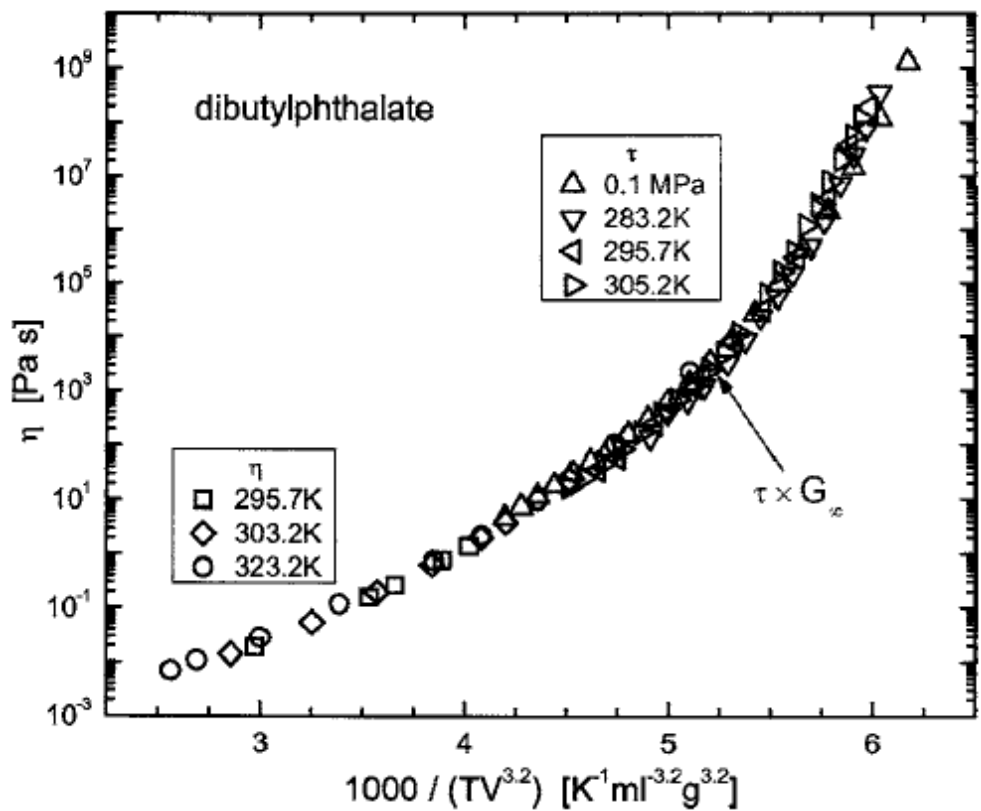

**Figure S5.** Superpositioned viscosities of dibutylphthalate with $\gamma=3.2$. The measurement pressure was from 0.1 to 1250 MPa at various temperatures as denoted by the symbol type. Dielectric relaxation times for $P \leq 1610$ MPa are included in the figure after multiplication by a factor of $2\times 108$ Pa. Figure reproduced by permission.

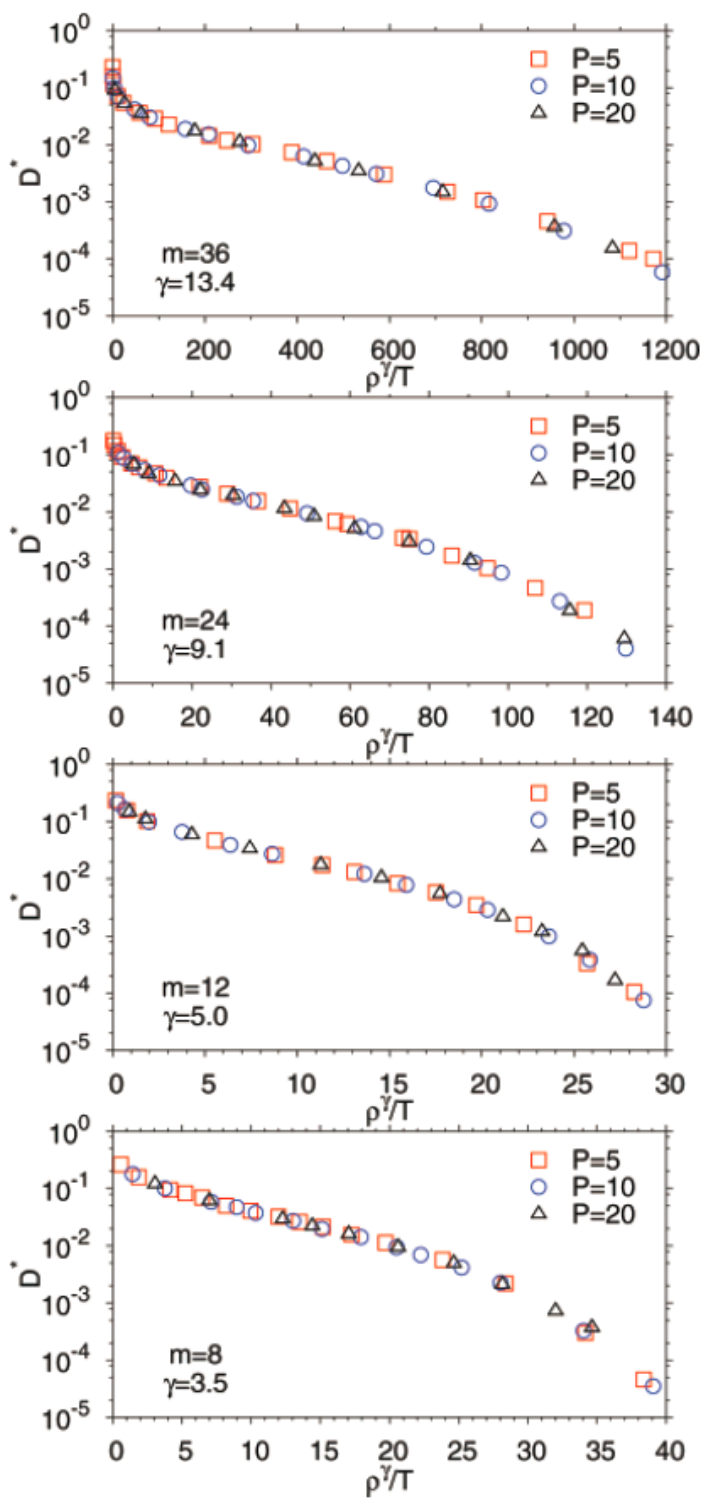

**Figure S6(a).** Reduced diffusion coefficients $D^*$ as a function of $F_\varsigma/T$ for different values of the repulsive exponent $m$ at different pressures: $P = 5$ (squares), $P = 10$ (circles), and $P = 20$ (triangles). From top to bottom: $m = 36$ ($\gamma=13.4$), $m = 24$ ($\gamma= 9.1$), $m = 12$ ($\gamma= 5.0$), and $m = 8$

($\gamma$= 3.5). The estimated uncertainty on $\gamma$ is ±0.1 (±0.2 for $m$ = 36). Figure reproduced by permission.

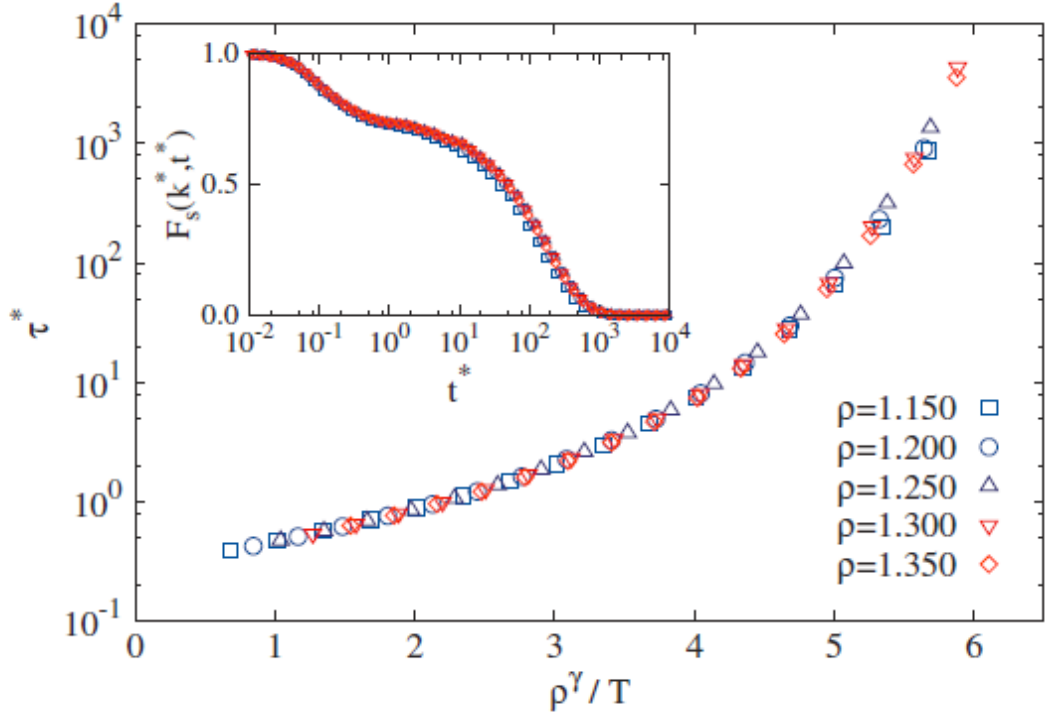

**Figure S6(b).** Reduced relaxation times $\tau^* = \tau(\rho^{1/3}T^{1/2})$ as a function of $\rho^\gamma/T$ with $\gamma = 5.1$ for all studied densities. Inset: Self-intermediate scattering functions as a function of reduced time $t^* = t(\rho^{1/3}T^{1/2})$ for state points at which $\rho^\gamma/T = 5.07$: $T=0.402$ (at $\rho=1.15$), $T=0.50$ (at $\rho=1.20$), $T=616$ (at $\rho=1.30$), and $T=0.912$ (at $\rho=1.350$). A constant reduced wave-vector $k^* = k(\rho^{1/3}) = 7.44$ is considered. Figure reproduced by permission.

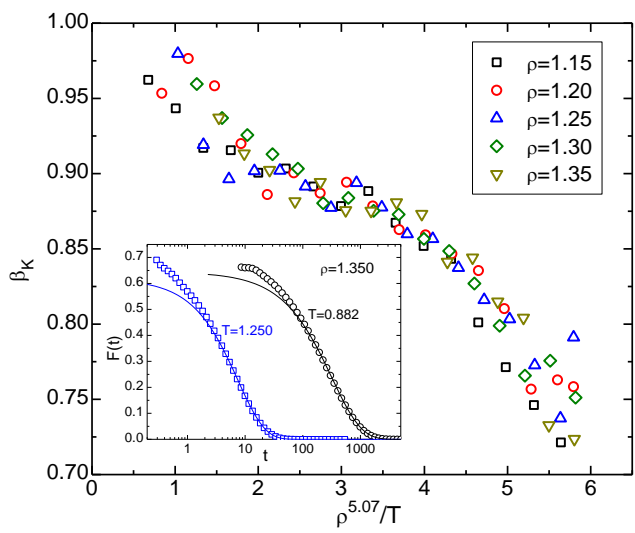

**Figure S6(c).** Kohlrausch exponent as a function of the product variable $\rho^{5.07}/T$ for simulated Lennard-Jones particles. Each symbol represents a distinct state point. Fits to the intermediate scattering function at longer time are shown for two temperatures in the inset. Figure reproduced by permission.

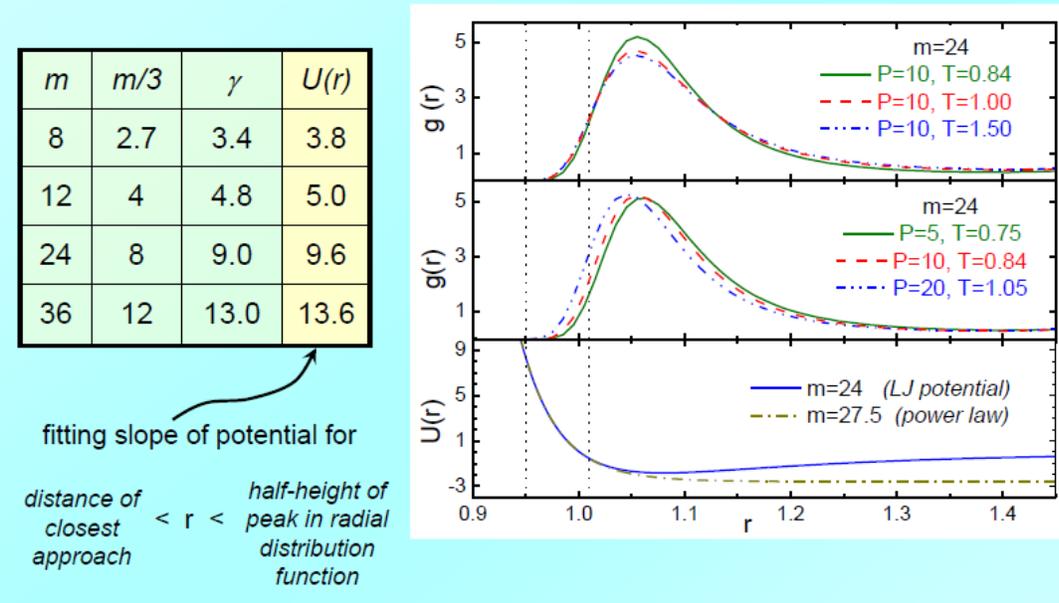

**Figure S7.** Top panel: radial distribution functions between large particles $g_{11}(r)$ at $P = 10$ for $T < T_O$: $T = 1.20$ (dotted), $T = 1.00$ (dashed), and $T = 0.84$ (solid). Middle panel: $g_{11}(r)$ at the lowest equilibrated $T$: $T = 0.75$ at $P = 5$ (dotted), $T = 0.84$ at $P = 10$ (dashed), and $T = 1.05$ at $P = 20$ (solid). Bottom panel: pair potential $U(r)$ (solid) and fitted IPL (dotted) in the range [0.95:1.01]. The latter range is indicated by vertical dotted lines in all panels. In the table comparisons are made for the scaling exponent $\gamma$ with $m/3$ and with the slope of the potential at short distance in the range as indicated. Figure courtesy of C.M. Roland.

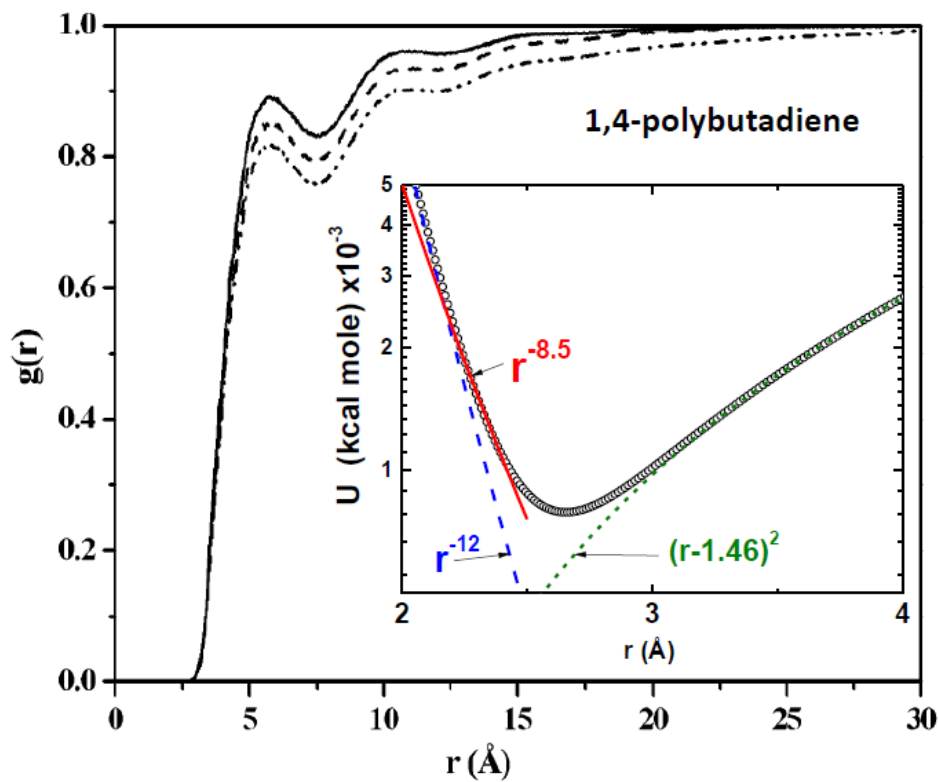

**Figure S8.** Intermolecular pair distribution function, $g(r)$, obtained from the MD simulations of cis-1,4-polybutadiene with three different chain lengths, the C48, C96, and C240 at 413 K, P = 1 atm by Tsolou et al.[53] The inset is the intermolecular potential for 1,4-polybutadiene calculated by Roland et al.[16] as the sum of an LJ 6–12 intermolecular potential and a harmonic intramolecular bond stretching potential with the parameters were taken from Ref. 53, with average values used for the interaction parameters. The solid line is the fit to the repulsive energy in the vicinity of the minimum, the long dashed line represents the limiting value of the repulsive energy at small $r$, and the short dashed line is the bond stretching interaction. Figure reproduced by permission.

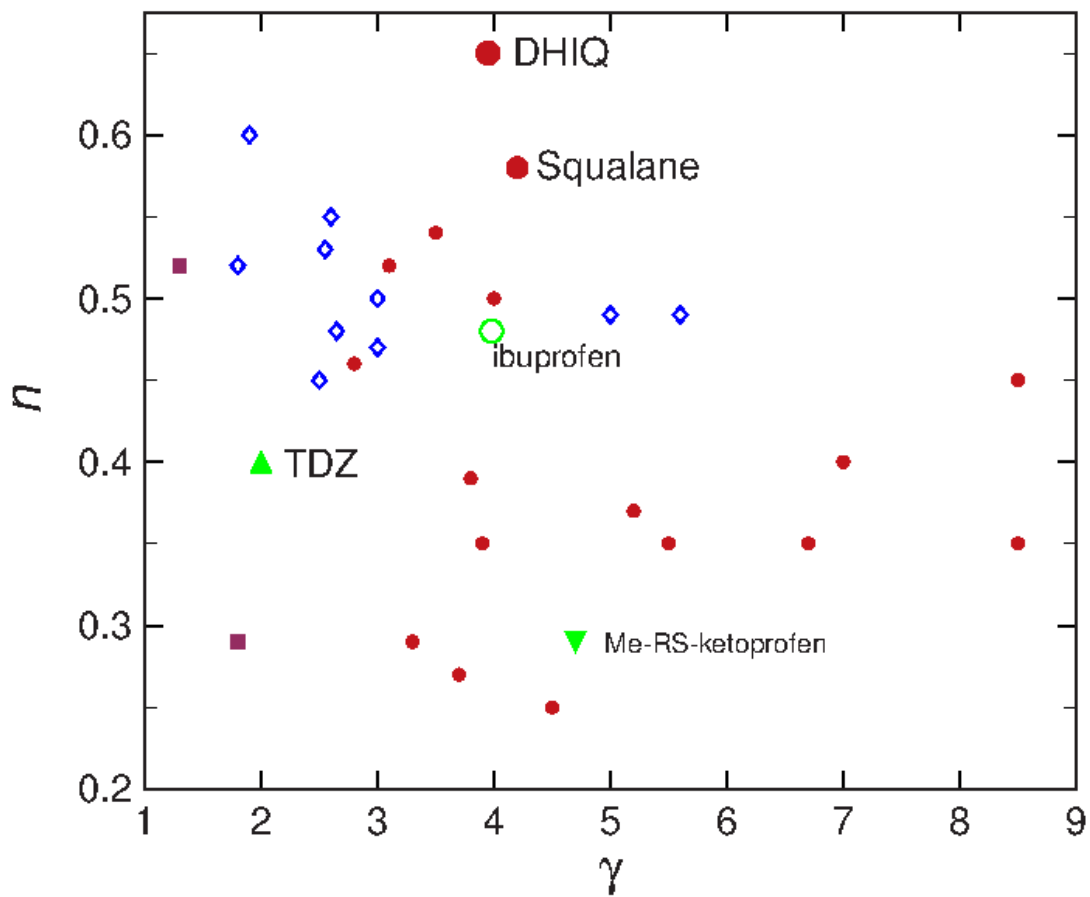

**Figure S9.** Plot of *n* against $\gamma$, where the closed circles are data of small molecular van der Waals glassformers, the open diamonds are polymeric glassformers, and the two closed squares are data of glycerol (smaller value) and sorbitol (larger value). Added are more recent data from DHIQ and squalene from Casalini and Roland, and TDZ, ibuprofen and Me-RS-ketoprofen discussed in this paper. There is no correlation between *n* and $\gamma$. For source of data see Ref.(40).

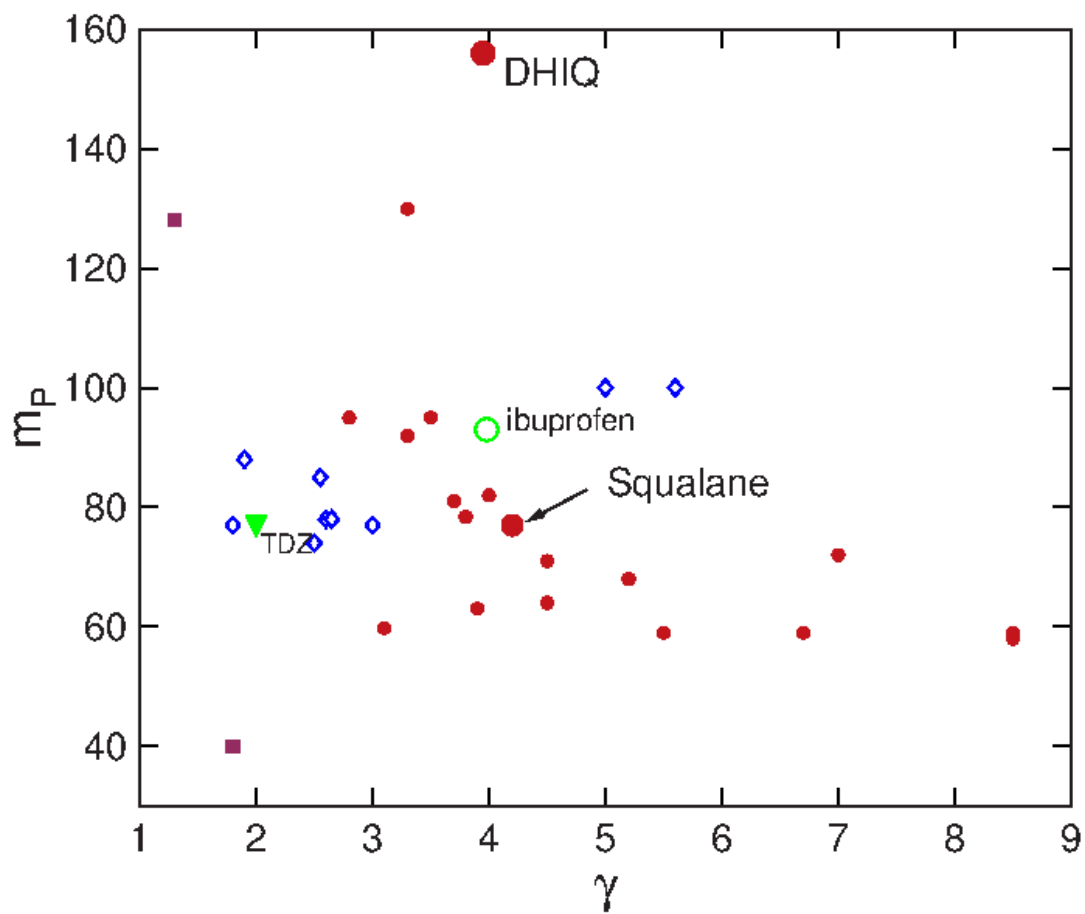

**Figure S10.** Plot of isobaric fragility index, $m_P$, against $\gamma$. The symbols stand for the same glassformers as in the left panel. There is no correlation between $m_P$ and $\gamma$. For source of data see Ref.(40).

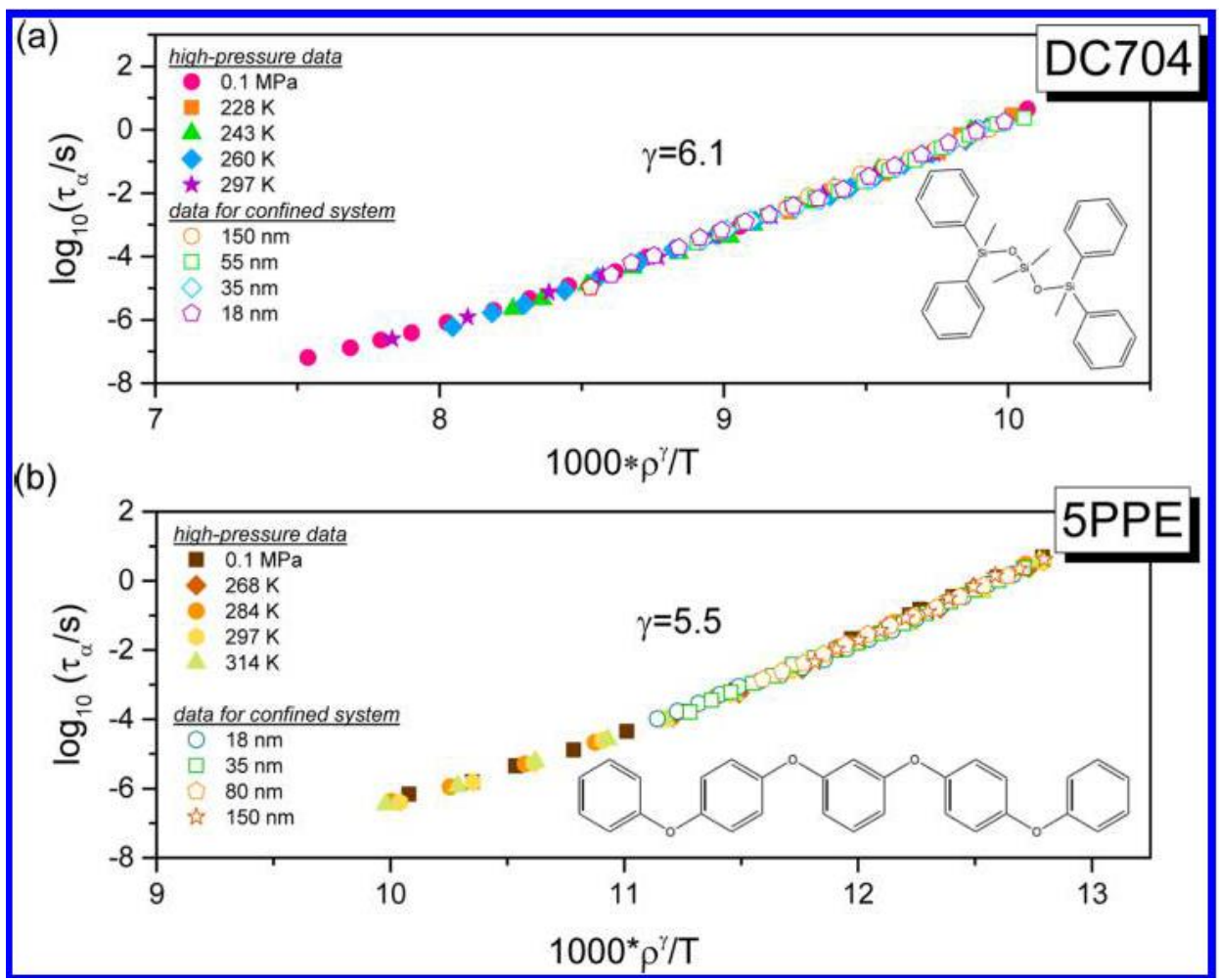

**Figure S11.** Dielectric relaxation time $\tau\alpha$ as a function of $\rho^\gamma/T$ with (a) $\gamma = 6.1$ and (b) $\gamma = 5.5$ for DC704 and 5PPE, respectively. Confinement data include $\tau_\alpha(T)$ dependences measured in the temperature range between $T_{g,core}$ and $T_{g,interface}$. Isobaric and isothermal data from the high-pressure studies are shown as well. Figure reproduced by permission.